\documentstyle[12pt,a4]{article}
\newtheorem{theorem}{Theorem}

\begin{document}

\begin{center}
{\huge \bf When are Vector Fields Hamiltonian?}\\
\vspace{1.5truecm}
{\large \bf P. CREHAN}\\
\vspace{1.5truecm}
Dept. of Mathematics, Faculty of Science,
University of Kyoto, Kyoto 606, Japan.\\
pat@kusm.kyoto-u.ac.jp
\end{center}
\vspace{2truecm}
There is reason to believe that at small scales and low temperatures, quantum
mechanical effects will play a role in dissipative systems which arise in
solid state physics. References to and a brief discussion of various
approaches to the quantisation of the damped harmonic oscillator,
can be found in \cite{crehan2}.
Although there exist successful and physically intuitive ways to deal with the
quantisation of the damped harmonic oscillator, for reasons outlined
in \cite{crehan2} we consider a more direct approach to the problem, one
which requires an understanding of the Hamiltonian  structure of  the
classical flow. This is why we ask the question from
which the article gets its title.

Locally it is possible to consider any flow to be Hamiltonian in
a sense which we will later make precise. This is true even if the vector
field is not conservative. Having understood the
Hamiltonian structure of the classical flow, it is possible in
principle to *-quantise the system by constructing an associated
Moyal-type algebra. This  step is based on an
insight of Bayen, Flato, Fronsdal Lichnerowicz and Sternheimer
\cite{bayenetal},
who pointed out that the Moyal algebra based on the product

$$
A(q,p)*B(q,p) = \exp\left( \imath {\hbar\over 2} \Lambda^{ij}
\partial_i^{(1)}\ \partial_j^{(2)} \right)
\ A^{(1)}(q,p)\ B^{(2)}(q,p)
$$

\noindent
where $\Lambda^{ij}$ is the standard symplectic matrix, is an associative
deformation of the classical commutative associative algebra of observables.
Although the Moyal algebra is widely used in the context of
the Wigner formalism, the general problem of
constructing Moyal-like algebras associated with a
given Hamiltonian structure is not understood, and there
remain many problems to solve before such a program for quantisation
can be completed.

The purpose of this talk is to highlight the fact that it is possible to
consider dissipative flows as autonomous Hamiltonian systems, to
clarify what this means, and to outline further problems that
this path to quantisation presents.
Our starting point is the following theorem \cite{crehan1}.

\begin{theorem} In a finite but sufficiently small
neighbourhood $U$ of a non-critical point of a smooth vector field
$v=v^i\partial_i$ on $\Re^{n+1}$, there exist independent
functions $\Phi^{(k)}\in C^2(U)$, a scalar $\rho\in C^1(\Re^{n+1})$ and smooth
independent rank 2
Poisson tensors $\Lambda_{(k)}$, $k=1,\ldots,n,$ such that

$$ v^i = \Lambda^{ij}_{(k)}\ \partial_j\Phi^{(k)},$$

\noindent
where

$$ \Lambda^{i j}_{(k)}=\rho^{-1}\ \epsilon^{i s_1\ldots s_{k-1} j
s_{k+1} \ldots s_{n}}\ \partial_{s_1}\Phi^{(1)}\ldots \partial_{s_{k-
1}}\Phi^{(k-1)}\
\partial_{s_{k+1}}\Phi^{(k+1)}\ldots\partial_{s_n}\Phi^{(n)}.$$
\end{theorem}

The $\Phi^{(k)}$ can be thought of as local Hamiltonians of the flow.
The Poisson tensors $\Lambda^{ij}_{(k)}$ are so called because each
operation defined by
$\{ A, B \}_{(k)} = \Lambda^{ij}_{(k)}\partial_i A\ \partial_j B$,
has all the algebraic properties of the familiar Poisson bracket.
In addition a local measure constructed from $\rho$, and given by
$d\mu=\rho dx^{0}\wedge\ldots\wedge dx^{n}$, is invariant under
the flow.

\section{The Kepler system}
First we consider a system which is Hamiltonian by all of the
usual criteria, the Kepler system on $(\Re^3-\{0\})\times\Re^3$.
This system
is exceptional even among fully integrable systems on a $2n$
dimensional phase space, since it has a maximum number of global
conservation laws $2n-1$, instead of $n$. These can be chosen from
the components of the angular momentum vector

\begin{eqnarray*}
L_1&=&y p_z-z p_y\\
L_2&=&z p_x-x p_z\\
L_3&=&x p_y-y p_x
\end{eqnarray*}

\noindent
and those of the Runge-Lenz vector

\begin{eqnarray*}
R_1&=&p_y L_3-p_z L_2-m x/r\\
R_2&=&p_z L_1-p_x L_3-m y/r\\
R_3&=&p_x L_2-p_y L_1-m z/r
\end{eqnarray*}

\noindent
where $r^2=x^2+y^2+z^2$, and $m$ is the mass.
Putting $\rho = - L_3 (L_1^2+L_2^2+L_3^2)$, $\Phi^{(2)}=L_2$,
$\Phi^{(3)}=L_3$, $\Phi^{(4)}=R_1$, and $\Phi^{(5)}=R_2$,
we can verify that

$$\Lambda^{ij}_{(1)}= \rho^{-1}\ \epsilon^{i j s_2 s_3 s_4 s_5}
\partial_{s_2}\Phi^{(2)}\ \partial_{s_3}\Phi^{(3)}\
\partial_{s_4}\Phi^{(4)}\ \partial_{s_5}\Phi^{(5)}$$

\noindent has all the properties required of a Poisson bracket. Taking
$\Phi^{(1)}=L_1$, to be the Hamiltonian, and computing
$\dot{x}^i=\Lambda^{ij}_{(1)}\partial_j
\Phi^{(1)}$ we recover the familiar Kepler flow on $\Re^6$, and we can
check that
$\rho\ dx^1\wedge \ldots \wedge dx^6$ does indeed provide an invariant
measure.
This is not the only way in which to construct the Kepler system.
In fact it is easy to show that this can be done
in an infinite number of different ways.

\section{The simple predator-prey equation}
For flows on $\Re^2,$ the picture is a little simpler. The Poisson tensors
all have the form $\Lambda^{ij}=\rho^{-1}\epsilon^{ij}$, where
$\rho\in C^1(\Re^2)$. Hamiltonian vector fields with respect to any such
Poisson structure, have the form $\dot{x}^i = \Lambda^{ij}\partial_j \Phi$.
The corresponding symplectic two-forms are given by $\rho\ dx\wedge dy$ and
these co-incide with the invariant measures mentioned above.

We will illustrate this with a system which is not usually
treated within the Hamiltonian framework,
the simple predator-prey equation. This system is not Hamiltonian
with respect to the canonical Hamiltonian structure on $\Re^2$, and
the flow does not conserve the standard volume element. On the other
hand it is not dissipative, and
it can be considered to be globally Hamiltonian with respect
to an infinite number of non-standard Hamiltonian structures.
It is defined by

\begin{eqnarray*}
\dot{x}&=& x(b-y) \\
\dot{y}&=& y(x-a)
\end{eqnarray*}

\noindent where $a,b\in \Re\ge 0.$ Taking

\begin{eqnarray*}
\Phi&=&\exp(x+y)\ x^{-a}\ y^{-b}\\
\rho&=&-\exp(x+y)\ x^{-(a+1)}\ y^{-(b+1)}\\
\end{eqnarray*}

\noindent
it is easy to check that the simple predator-prey system is
Hamiltonian with
Hamiltonian $\Phi$, with respect to the Poisson structure
$\Lambda^{ij}=\rho^{-1}\epsilon^{ij}$. To see that it
has an infinite number of other Hamiltonian structures, it suffices
to note that replacing
$\Phi$ with any non-constant $F(\Phi)$, and $\rho$ with
$\rho F^{\prime}(\Phi)$ we get a different Hamiltonian function and a
different Poisson structure, for the same system.

$\Phi$ is well behaved except when $x=0$ or $y=0$, and $\Lambda$ becomes
singular along these lines. We know that
the flow never crosses these axes, so it is globally Hamiltonian
on $(0,\infty]\times (0,\infty]$.

\section{The damped harmonic oscillator}
Now we come to one of the simplest dissipative systems,
the damped harmonic oscillator. This is given by

\begin{eqnarray*}
\dot{x}&=&y         \\
\dot{y}&=&-x-\kappa y
\end{eqnarray*}

\noindent where $\kappa$ is the damping constant. The
qualitatively different phases - undamped (H),
damped (D), critically damped (C), and over-damped (O), correspond to
$\kappa=0,$ $0<\kappa<2,$ $\kappa=2$ and $\kappa>2$
respectively. For each phase it is possible to determine an infinite number
of Hamiltonian structures. However the following choice has
the property that as $\kappa$ descends from high values down through the
value $2$ to $\kappa=0$, the Hamiltonians vary continuously from
$\Phi^O$ to $\Phi^C$, from $\Phi^C$ to $\Phi^D$ eventually to
$\Phi^H$ at $\kappa=0$ where we recover the familiar case of the
harmonic oscillator.

\begin{eqnarray*}
\Phi^H&=&{1\over 2}(x^2+y^2)\\
\Phi^D&=&{1\over 2}(x^2+y^2+\kappa x y)\exp\left(2{\kappa\over \Delta}
\arctan\left({{\Delta x}\over {\kappa x + 2y}}\right)\right)\\
\Phi^C&=&{1\over 2}(x+y)^2 \exp\left({{2x}\over{x+y}}\right)\\
\Phi^O&=&{1\over 2}(x^2+y^2 + \kappa x y) \left|{{\kappa x + 2 y
+\Delta x}\over {\kappa x + 2 y - \Delta x}} \right|^{\kappa\over
\Delta}
\end{eqnarray*}

\noindent For damped motion $\Delta = +(4 - \kappa^2)^{1/2}$ and
for overdamped motion $\Delta = +(\kappa^2 - 4)^{1/2}$. On $\Re^2$
the invariant volume element provided by $\rho\ dx\wedge dy$
is also the symplectic
structure corresponding to the relevant Poisson structure.

\begin{eqnarray*}
\rho^H&=&1\\
\rho^D&=&\exp\left(2{\kappa\over \Delta}
\arctan\left({{\Delta x}\over {\kappa x + 2 y}}\right)\right)\\
\rho^C&=&\exp\left({{2x}\over{x+y}}\right)\\
\rho^O&=&\left|{{\kappa x + 2 y + \Delta x}\over
{\kappa x + 2y - \Delta x}}\right|^{\kappa \over \Delta}.
\end{eqnarray*}

In this way we can consider the Hamiltonians and Poisson structures
of the damped harmonic oscillator as one-parameter
deformations of the simple harmonic oscillator. In \cite{crehan2}
we deal with this in more detail and
relate the different phases of the flow to the nature
of the singularity in the various structures.

\section{Discussion}
We have shown by examples how many different kinds of vector fields
can be considered to be Hamiltonian. These three examples
present different kinds of problems
for the *-quantisation program. The Kepler system is well understood
within the standard framework of Hamiltonian dynamics
on $\Re^3\times\Re^3$, however the existence
of many different Hamiltonian structures
raises the question of whether these will give rise to many
different quantisations.
The predator-prey system was shown to be globally Hamiltonian on the upper
right quadrant of the plane. In this case the usual division of
phase space into position and momentum does not readily
make sense. Both co-ordinates have the same physical interpretation
as representing the size of the corresponding population. From this point of
view an explicit phase-space quantisation seems more natural. However
none of its Hamiltonian structures
co-incide with the standard one on $\Re^2$, and a-priori we have no way of
choosing which one is correct.
The Poisson structures which arise in the case of the
damped harmonic oscillator are either multi-valued or singular.
This leaves the problem of boundary conditions for the wave-function along the
singularity. We think that this is a physically reasonable consideration
as singularities must also occur in conserved measures in
the classical case.
If eventually, we  succeed in consistently applying the *-quantisation
approach to the damped harmonic oscillator, it will be interesting to
see if there occur qualitatively different phases, which correspond
with those of the classical flow.

As an approach to the quantisation of general dynamical
systems, this program raises more questions than it answers.
However we think that these are questions which should be asked anyway of
standard quantum mechanics, and that the answers will be of
mathematical and physical interest in their own right.


\begin{thebibliography}{99}
\bibitem{crehan2} P. Crehan: Preprint Kyoto-Math 94-04: On the Hamiltonian
structure of non-conservative flows on $\Re^2$.
\bibitem{bayenetal} F. Bayen, M. Flato, C. Fronsdal, A. Lichnerowicz \&
D. Sternheimer: Ann. Physics (NY) vol.111 1978 p.61, p.111.
\bibitem{crehan1} P. Crehan: Preprint Kyoto-Math 94-03: On the local
Hamiltonian
structure of vector fields. To appear in Modern Physics Letters A.
\end{thebibliography}
\end{document}